\documentclass{PoS}

\usepackage{graphicx}
\usepackage{amsmath}
\usepackage{subfigure}
\usepackage{wrapfig}
\usepackage{epsfig}

\newcommand{\be}{\begin{equation}}
\newcommand{\ee}{\end{equation}}
\newcommand{\bea}{\begin{eqnarray}}
\newcommand{\eea}{\end{eqnarray}}

\def\bea{\begin{eqnarray}}
\def\eea{\end{eqnarray}}

\def\lmatrix{\left(\begin{array}}
\def\rmatrix{\end{array}\right)}

\title{Nearly conformal electroweak sector \\ with chiral fermions}

\ShortTitle{Nearly conformal electroweak sector with chiral fermions}

\author{Zolt\'an Fodor\\
        Department of Physics, University of Wuppertal\\
        Gau$\beta$strasse 20, D-42119, Germany\\
        Email: \email{fodor@bodri.elte.hu}}

\author{Kieran Holland\\
        Department of Physics, University of the Pacific\\
        3601 Pacific Ave, Stockton CA 95211, USA\\
        Email: \email{kholland@pacific.edu}}

\author{Julius Kuti \\
        Department of Physics 0319, University of California, San Diego\\
        9500 Gilman Drive, La Jolla, CA 92093, USA\\
        E-mail: \email{jkuti@ucsd.edu}}

\author{\speaker{D\'aniel N\'ogr\'adi}\\
        Department of Physics 0319, University of California, San Diego\\
        9500 Gilman Drive, La Jolla, CA 92093, USA\\
        Email: \email{nogradi@lorentz.leidenuniv.nl}}

\author{Chris Schroeder\\
        Department of Physics 0319, University of California, San Diego\\
        9500 Gilman Drive, La Jolla, CA 92093, USA\\
        E-mail: \email{crs@physics.ucsd.edu}}

\abstract{
$SU(3)$ gauge theory with dynamical overlap fermions in the 2-index symmetric (sextet)
representation is considered. This model may be a viable model of the electroweak symmetry breaking
sector along the lines of the walking technicolor paradigm. The number of fermion species is chosen
such that the theory is expected to be below the conformal window. We will discuss how the $\varepsilon$-regime
and random matrix theory can be used to test whether at any given set of parameters ($N_c$, $N_f$, representation)
the theory is in the conformal phase or indeed just below it. Quenched Monte Carlo results are included in the fundamental
representation and also preliminary dynamical ones in the 2-index symmetric representation.
}

\FullConference{The XXVI International Symposium on Lattice Field Theory\\
		 July 14 -- 19 2008\\
		 Williamsburg, VA, USA}

\begin{document}

\section{Introduction}

Model building of a strongly interacting electroweak sector, with or without a Higgs resonance,
requires the knowledge of the phase diagram of non-abelian gauge theories for varying number of colors $N_c$, number of
fermion flavors $N_f$, and representation $R$. For fixed $N_c$ and $R$ the theory is generically in the chiral 
symmetry broken phase for low $N_f$ and the conformal phase for high $N_f$ as long as asymptotic freedom is maintained,
i.e. $N_f$ is not too high. Certain models require $N_f$ to be just below the conformal window along the lines of the
walking technicolor paradigm \cite{Weinberg:1975gm} and the knowledge of the critical $N_f$ separating the two phases is essential.

Mapping out the phase diagram in the space of $N_c$, $N_f$ and $R$ is an interesting problem on its own and can be
useful for model builders with different motivations such as unparticles. We are first and foremost concerned with the Higgs
mechanism though.

In this context the parameters $N_c$, $N_f$ and $R$ are not only restricted by the phase diagram but also by electroweak precision data
and the symmetry breaking pattern necessary for generating masses for the $W$ and $Z$ bosons. Consistency with electroweak
precision data requires a small $S$-parameter while the simplest symmetry breaking pattern is the one which generates exactly
3 Goldstone bosons with no (techni)pions left over after the massive gauge bosons acquired their masses. The simplest model
fulfilling these requirements is $SU(3)$ gauge group with $N_f = 2$ fermions in the 2-index symmetric ($2S$) representation which
is the topic of our study.

% So far the most widely studied technicolor models included fermions in the fundamental representation of $SU(3)$ which choice combined with
% the requirement of a walking coupling constant necessitates a relatively large $N_f$ resulting in a presumably large $S$-parameter.
% Nevertheless the methods necessary to study nearly conformal theories can be put to a test in this case too.

In a numerical simulation at finite volume, finite lattice spacing and (usually) finite quark masses it is a non-trivial task to
determine whether the theory is conformal in the continuum, massless quark and infinite volume limits, or chiral symmetry is
broken just as in QCD. In section \ref{diracspectrum} a method is introduced that is capable of distinguishing the two phases
based on the behavior of the low-lying Dirac eigenvalues. If the theory is QCD-like chiral symmetry is spontaneously broken and
random matrix theory (RMT) will predict the distribution of low-lying eigenvalues in the $\varepsilon$-regime \cite{Gasser:1987ah}, whereas in the conformal
phase chiral symmetry is unbroken and the spectral density of the Dirac operator goes to zero around $\lambda = 0$. One
particular advantage of RMT is that it works for finite (but small of course) quark mass. This method
of distinguishing phases with the help of the Dirac spectrum has been applied for dynamical staggered fermions in \cite{kieran} which
complements Schrodinger functional and finite temperature based investigations of similar dynamical staggered
models \cite{Appelquist:2007hu, Deuzeman:2008sc}.

\section{Perturbative expectations}

As is well-known the 2-loop perturbative $\beta$-function \cite{Caswell:1974gg} can be used to estimate the critical $N_f$ value above which the theory is
conformal and below which chiral symmetry is spontaneously broken \cite{Banks:1981nn}. If the first two coefficients are $\beta_1$
and $\beta_2$, 
% For $N_c$ colors, $N_f$ flavors of fermions
% in the irreducible representation $R$ the 2-loop $\beta$-function is
% \bea
% \beta(g) &=& - \left( \beta_1 \frac{g^3}{16\pi^2} + \beta_2 \frac{g^5}{(16\pi^2)^2} \right) \\
% \beta_1 = \frac{11}{3} N_c - \frac{4}{3} T(R) N_f \;, &\qquad&
% \beta_2 = \frac{34}{3} N_c^2 - \left( \frac{20}{3} N_c + 4 C_2(R) \right) T(R) N_f \nn
% \eea
% where  $C_2(R)$ is the quadratic Casimir in $R$ and $T(R) (N_c^2 - 1) = C_2(R) \dim(R)$. 
asymptotic freedom requires $\beta_1 < 0$ 
% which leads to $N_f < N_f^{asympt} = 11 N_c/(4 T(R))$ 
otherwise the theory is free in the continuum.
If $\beta_1 < 0$ and $\beta_2 > 0$ a
non-trivial zero of the $\beta$-function exists hinting at a non-trivial IR fixed point.
% This is the case if $N_f > 17 N_c^2 / ( T(R) ( 10N_c + 6C_2(R) ))$.
% \bea
% \label{bz}
% N_f > \frac{17 N_c^2}{ T(R) ( 10N_c + 6C_2(R) )}\;.
% \eea
However if the fixed point value of
the coupling is too large chiral symmetry is spontaneously broken before the flow in the IR can reach
the would-be fixed point. 
% Since the value of the coupling constant at the zero of the $\beta$-function 
% decreases with increasing $N_f$ and is certainly very small just below the loss of asymptotic freedom
It is nevertheless expected that a critical $N_f^{crit}$ value exists above which the theory is really conformal even non-perturbatively. 
% In the conformal window $N_f^{crit} < N_f < N_f^{asympt}$ chiral symmetry is not broken and the spectrum is gapless.

The value $N_f^{crit}$ can be estimated in the ladder approximation by the requirement that the anomalous dimension of ${\bar\psi}
\psi$ reaches $\gamma = 1$ \cite{Appelquist:1988yc}. Using this bound,
% The result from this analysis is $N_f^{crit} = ( 17N_c + 66 C_2(R)) / ( T(R)(10N_c+30C_2(R)))$.
% \bea
% N_f^{crit} = \frac{17N_c + 66 C_2(R)}{T(R)(10N_c+30C_2(R))}\;.
% \eea
the conformal window for $SU(2)$ and representations $j=1/2, 1$ and $3/2$ is expected to be $8 < N_f < 11$,
none (for integer $N_f$) and $N_f = 1$ respectively, and no window for $j>3/2$.
For $SU(3)$ and fundamental, adjoint = $2A$ and $2S$ representations the conformal window is expected
to be $12 < N_f < 16$, none (for integer $N_f$) and $N_f = 3$ respectively.

\section{Dirac spectrum}
\label{diracspectrum}

To what extent the perturbative expectations of the previous section are justified is an open question in general.
Non-perturbative tests of these expectations have been performed for various gauge groups, flavor number and representations
\cite{ Appelquist:2007hu, Deuzeman:2008sc,Catterall:2007yx, Catterall:2008qk,  Shamir:2008pb, DeGrand:2008dh, Svetitsky:2008bw}
using various methods.

The low-lying spectrum of the Dirac operator is sensitive to the IR dynamics of the theory and shows characteristically
different behavior in the conformal and QCD-like phases. Its measurement is in principle straightforward in a lattice 
simulation hence it is a good candidate to distinguish the two phases.

\subsection{Chirally broken phase, $\varepsilon$-regime, random matrix theory}
\label{chirallybrokenphase}

If chiral symmetry is spontaneously broken, the Banks-Casher relation connects the spectral density $\rho(\lambda)$ of the Dirac operator
around zero to the chiral condensate \cite{Banks:1979yr},
\bea
\Sigma = \lim_{m\to0} \lim_{V\to\infty} \frac{\pi \rho(0)}{V}\;.
\eea
It also implies that the low-lying eigenvalues are dense in the sense that the average spacing is inversely proportional to the
volume,
\bea
\label{chideltal}
\Delta \lambda = \frac{\pi}{\Sigma V}\;.
\eea

It has been suggested long ago that if the bare parameters $\beta$, $m$
are tuned to the $\varepsilon$-regime, i.e. such that $m_\pi < L^{-1} < f_\pi$ the low-lying Dirac spectrum follows the predictions of a random matrix theory
\cite{Shuryak:1992pi, Verbaarschot:1994qf}.
The corresponding random matrix model is only sensitive to the pattern of chiral symmetry breaking, the topological charge and the rescaled fermion mass
once the eigenvalues are also rescaled by the same factor $\Sigma V$. 
% This suggestion has been confirmed in various settings both
% in quenched and fully dynamical simulations [ref].

More precisely, random matrix theory provides analytic formulae for the microscopic spectral density
\bea
\rho_S(\zeta) = \frac{1}{\Sigma V} \rho\left( \frac{\zeta}{\Sigma V} \right) = \sum_{k=0}^\infty p_k( \zeta )\;,
\eea
and the individual eigenvalue distributions $p_k(\zeta)$ where $\zeta = \lambda \Sigma V$. The distributions $p_k(\zeta)$
only depend on $\mu = m \Sigma V$, $N_f$ and the topological charge $\nu$. The value of $\Sigma$
can be obtained by using $\mu$ as the fitting parameter to have
\bea
\label{sigmafit}
\frac{\langle \zeta_k \rangle }{\mu} = \frac{\langle \lambda_k \rangle}{m}\;,
\eea
where the left hand side is calculated in random matrix theory at a fixed charge $\nu$ while the right hand side is measured in
the simulation in the given sector $\nu$. Which eigenvalue $\lambda_k$ and which sector $\nu$
is used is arbitrary in principle (as long as $k$ is not too large, say, $k=1,2,3$)
and the quality of the whole procedure may be characterized by the (in)consistency of the obtained $\Sigma = \mu / (mV)$
values for various $k$ and/or $\nu$. 

A more stringent test is the comparison of $p_k(\zeta)$ between the random matrix theory predictions and the simulation
once a consistent $\Sigma$ and corresponding $\mu$ have been obtained from the above fitting procedure. The agreement is
only expected for the first few eigenvalues.

In order to see the effects of dynamical quarks the lowest eigenvalue should be larger than the fermion mass $m$. Otherwise the
simulation is effectively quenched and random matrix theory will only agree at $N_f = 0$.

Out of the two requirements of the $\varepsilon$-regime, $m_\pi L < 1$ can be satisfied by tuning the fermion mass to a small
value at any $L$. However the second requirement, $f_\pi L > 1$, is largely independent of $m$ provided it is small enough
and puts a lower bound on $L$. As the lower edge of the conformal window is approached from below, $f_\pi$ is expected to decrease and
eventually will vanish as the theory becomes conformal. Hence the $f_\pi L > 1$ condition will be more and more difficult to satisfy
and larger and larger lattices will be needed the closer the theory is to the conformal window. As a result the study of
nearly conformal (or walking) technicolor models is very challenging in the $\varepsilon$-regime.

It should be noted that the requirement $f_\pi L > 1$ is valid up to numerical constants only. From the behavior
of the rotator and Goldstone spectrum of the chiral Lagrangian it can be made more precise as $f_\pi L > 1/\sqrt{2\pi}$
which is the requirement of these two spectra to separate from each other.
In fact we will see that in some cases RMT gives a good description even if $f_\pi L < 1$ which is probably due to the above
numerical constant $1/\sqrt{2\pi} = 0.3989...$ being smaller than 1.

\subsection{Conformal phase}

In the conformal phase no scale is generated and $\Sigma = 0$. The spectral density of the Dirac operator around $\lambda \sim 0$ behaves as
\bea
\label{confrho}
\rho(\lambda) \sim \lambda^{3+\gamma}
\eea
for massless quarks in the continuum and at infinite volume. Here $\gamma$ is the anomalous dimension of ${\bar\psi} \psi$.

The exact dependence of $\gamma$ on the conformal fixed point coupling $g_*$ is in 
principle calculable in perturbation theory since $g_*$ 
is presumably not large (otherwise chiral symmetry would be spontaneously broken). Certainly, $\gamma \sim g_*^2$. Of course only a non-perturbative
treatment can decide whether there is room for a fixed point coupling which is
large enough to be significantly different from perturbation theory and small enough so that chiral symmetry is not broken.

How the (\ref{confrho}) behavior is modified by finite volume and finite quark mass is an open question that we hope to address in the
future. Certainly, in the free $g_*=0$ case the average eigenvalue spacing is inversely proportional to the linear size $L$ of the box.
The characteristic feature that is expected to hold even for a finite $g_*>0$ is that the average eigenvalue spacing for small
eigenvalues is much less dense than in the chirally broken case where it is inversely proportional to the 4-volume $V$; see (\ref{chideltal}).

If the volume is too small, the chiral condensate is squeezed out of the box and the theory behaves perturbatively
even in the case when chiral symmetry is broken in an infinite volume. Hence great care is needed not to confuse a small volume
chiral symmetry breaking and a (small or large volume) conformal theory which is also behaving more-or-less perturbatively.

\section{Our model, $SU(3)$ with $N_f = 2$ in $2S$ representation}

The simplest example of a model that -- according to the perturbative expectations -- is just below the conformal window,
has a relatively low $N_f$ value so that the $S$-parameter is relatively small,
and has precisely 3 Goldstone bosons is $N_c = 3$, $N_f = 2$ and $R = 2S$.
This model has been studied in \cite{Shamir:2008pb, DeGrand:2008dh, Svetitsky:2008bw} using Wilson fermions on rather small lattices and it was found to be
already in the conformal window although it was indicated that more complicated possibilites are also allowed by the data.

Since exact chiral symmetry is important both for QCD-like and conformal theories we chose to use overlap fermions
\cite{Neuberger:1997fp}.
The simulation has to be carried out at a fixed topological charge. There are two methods available for simulating dynamical 
overlap fermions at fixed topology. One is the reflection/refraction algorithm \cite{Fodor:2003bh} but always reflecting on the topological
boundary. The other is employing a pair of extra Wilson fermions to suppress exact zero modes thereby
suppressing tunnelling between sectors \cite{Fukaya:2006vs}. We used the second method, which is much faster, in this study.

\section{Preliminary results}

\subsection{Quenched simulations}

In order to see how well actual simulations agree or disagree with the predictions of random matrix theory (RMT) we have tested the
RMT predictions for $N_f = 0$ and overlap valence quarks in the fundamental representation since in the quenched approximation
chiral symmetry is guaranteed to be broken. This setup is identical to \cite{Giusti:2003gf}
but actual eigenvalue distributions were not presented there. Since valuable information can be gained from these we decided to
redo this analysis on $12^4$ lattices at $\beta = 5.8458$ which corresponds to a lattice size of $L=1.49$ fm. All our parameters
were the same as in \cite{Giusti:2003gf}. In particular we used the Wilson gauge action and an unsmeared overlap operator.
Our ensemble consists of 1500 configurations.
The results of \cite{Giusti:2003gf} for expectation value ratios $\langle\lambda_i\rangle / \langle\lambda_j\rangle$
have been reproduced within 1-sigma precision with occasional 1.2-sigma deviations. 

\begin{figure}
\begin{center}
\begin{tabular}{ccc}
\includegraphics[height=3.2cm]{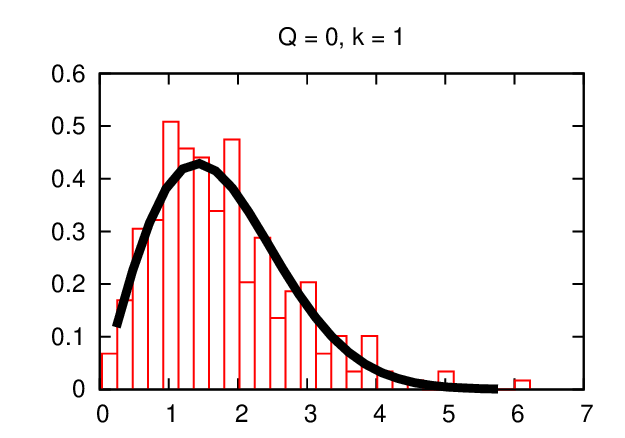}&\includegraphics[height=3.2cm]{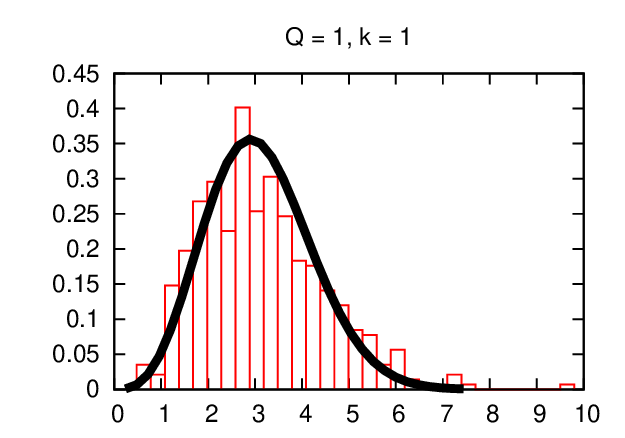}&\includegraphics[height=3.2cm]{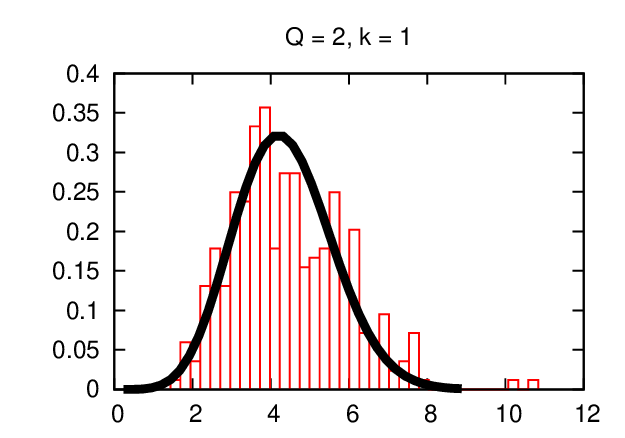}\\
\includegraphics[height=3.2cm]{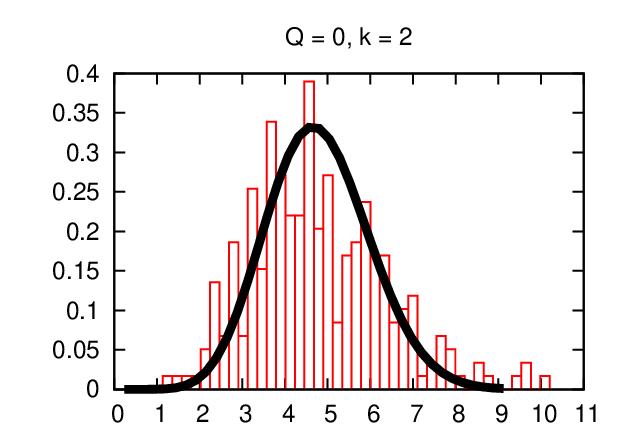}&\includegraphics[height=3.2cm]{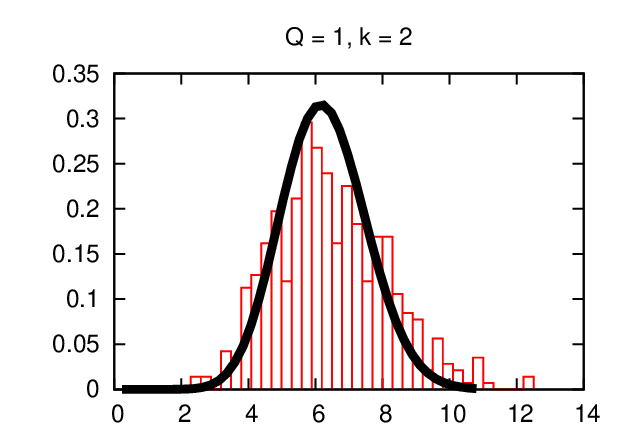}&\includegraphics[height=3.2cm]{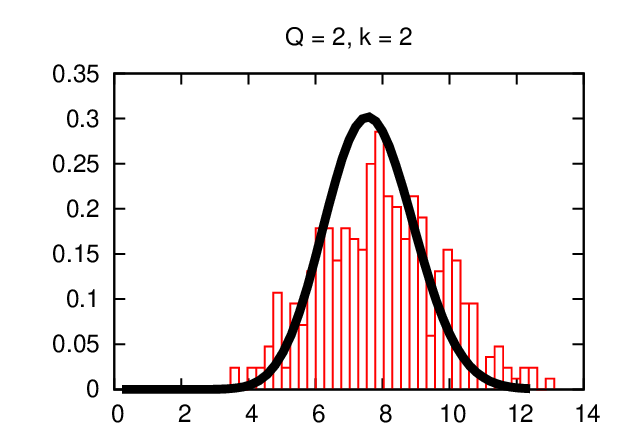}
\end{tabular}
\caption{Rescaled quenched eigenvalue distributions $p_k(\zeta)$ in the fundamental representation and $N_f = 0$ RMT predictions for $|Q| = 0, 1, 2$ and $k = 1, 2$\label{quenchedplots}}
\end{center}
\end{figure}

The distribution of the $k$th eigenvalue in various topological sectors $|Q|$ are shown in figure \ref{quenchedplots} together with
the RMT predictions once $\Sigma$ has been fitted from $k = 1$ and $Q = 0$ as described by (\ref{sigmafit}). Clearly, not only
the expectation values $\langle\lambda_i\rangle$ follow the RMT predictions but also their width. This level of agreement with RMT was 
not expected since $f_\pi L < 1$ for this ensemble, however as indicated in section \ref{chirallybrokenphase} a more accurate
requirement of the $\varepsilon$-regime is $f_\pi L > 1/\sqrt{2\pi}$ which this ensemble does fulfill.

A similar comparison for the 2-index symmetric representation in the quenched approximation is ongoing. Since RMT is only
sensitive to the topological charge, $N_f$, $\mu$ and the pattern of chiral symmetry breaking, the same random matrix model is expected
to describe this representation as the one used for the fundamental. Agreement with the same RMT using a different representation
than fundamental will be a non-trivial check of its universality. 
% A good starting point for our
% quenched study is \cite{Kogut:1984sb}
% where $T_c$ was measured in the quenched approximation with staggered quarks in the fundamental,
% $2S$ and $2A$
% representations of $SU(3)$. The shift in $\beta_c$ for the $2S$ representation relative to the fundamental is quite
% large, going from about $\beta_c = 5.7$ to $\beta_c = 7.8$ (using the Wilson plaquette action).

\subsection{Dynamical simulations}

In the 2-index symmetric representation three dynamical ensembles were generated on $6^4$ lattices
using the tree-level improved Symanzik gauge action at $\beta =
4.850,\; 4.975$ and $5.100$ and $N_f = 2$ flavors of massive quarks with $m = 0.05$. The negative Wilson mass in the overlap
operator was $m_W = -1.3$ and 2 levels of stout smearing with smearing parameter $\rho = 0.15$ have been applied. The 
topology change suppressing action of \cite{Fukaya:2006vs} was used with mass $M = 0.2$ for the ghost Wilson fermions and only the topological sector $Q = 0$ was
sampled. Since chiral symmetry is preserved by overlap fermions at finite lattice spacing $N_f = 2$ RMT is applicable. Fitting
$\Sigma$ from the average first eigenvalue as in (\ref{sigmafit}) one obtains $0.083(4)$, $0.084(4)$ and $0.080(4)$ in lattice
units for the three $\beta$ values respectively. 

\begin{figure}
\begin{center}
\begin{tabular}{ccc}
\includegraphics[height=3.2cm]{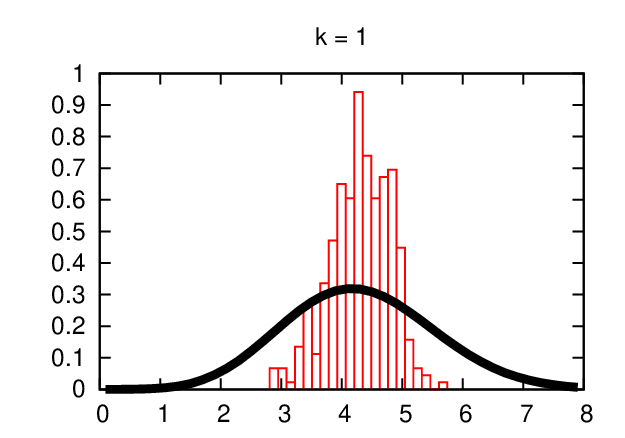}&\includegraphics[height=3.2cm]{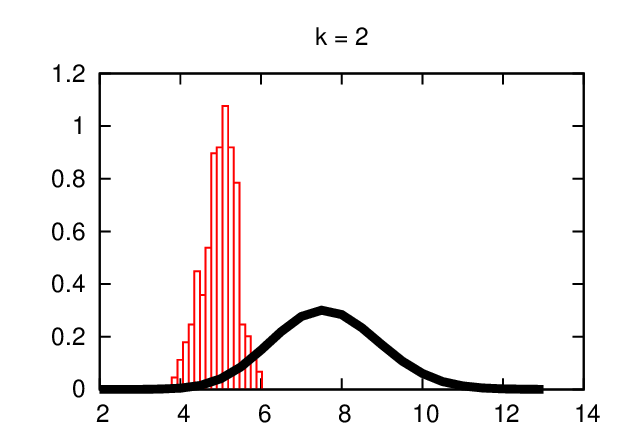}&\includegraphics[height=3.2cm]{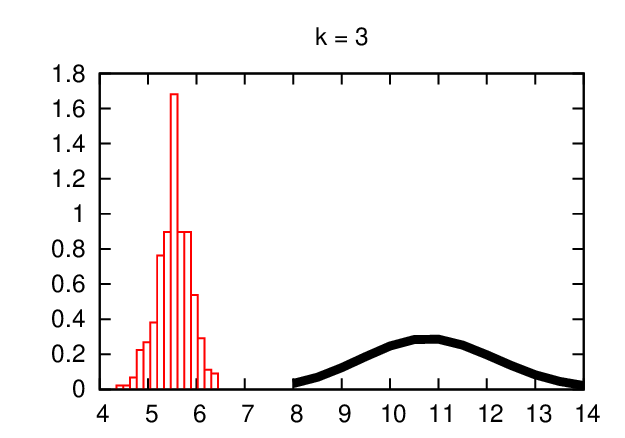}
\end{tabular}
\caption{Rescaled dynamical eigenvalue distributions $p_k(\zeta)$ in the 2-index symmetric representation and $N_f = 2$ RMT
predictions for $Q = 0$ and $k = 1, 2, 3$\label{2symplots}}
\end{center}
\end{figure}

The eigenvalue distributions look qualitatively the same for the three $\beta$ values and the first 3 eigenvalues are plotted in
figure \ref{2symplots} for $\beta = 4.850$ together with the $N_f = 2$ RMT predictions after rescaling both the eigenvalues and the quark
mass.

Clearly, the RMT predictions are very far from the simulation results, neither the averages nor the widths follow the RMT curves.
This may be due to several reasons the most likely of which is small volume. We have not measured either $m_\pi$ or $f_\pi$ so it
is not clear if the simulation was in the $\varepsilon$-regime at all. Simulations on larger volumes as well as measurements of
$m_\pi$ and $f_\pi$ are ongoing.

If the larger volume simulations agree with the above conclusion the 2-index symmetric representation for gauge group $SU(3)$
and $N_f = 2$ is already in the conformal window.

\section{Conclusions and outlook}

Needless to say that the results on the 2-index symmetric representation are preliminary. The quenched fundamental
representation simulations shows that the RMT predictions are very precise for the first few eigenvalues once the volume is large
enough. A similar conclusion is expected for the quenched $2S$ representation which will in addition
test the universality of RMT. Presently, the result of the $6^4$ dynamical $2S$ representation simulation are preliminary
and the deviation from RMT is thought to be due to small volume.
Larger volume simulations are ongoing for both the quenched and fully dynamical cases.

\subsection*{Acknowledgements}
D.N. would like to acknowledge helpful discussions with Christian Hoelbling, Tam\'as Kov\'acs and K\'alm\'an Szab\'o.
This research was supported by the DOE under grants DOE-FG03-97ER40546,
DE-FG02-97ER25308, by the NSF under grant 0704171, by DFG under grant FO 502/1, 
and by the SFB under grant SFB-TR/55.

\end{document}